# Ion-dependent dynamics of DNA ejections for bacteriophage lambda

Authors:
David Wu[1]*, David Van Valen[1]*, Qicong Hu[2], Rob Phillips[1]
1. Division of Engineering and Applied Sciences, California Institute of Technology, Pasadena, California.
2. School of Medicine, Stanford University, Palo Alto, California.
* These authors contributed equally to this work.

## Abstract

We study the control parameters that govern the dynamics of *in vitro* DNA ejection in bacteriophage lambda. Past work has demonstrated that bacteriophage DNA is highly pressurized; this pressure has been hypothesized to help drive DNA ejection. Ions influence this process by screening charges on DNA; however, a systematic variation of salt concentrations to explore these effects has not been undertaken. To study the nature of the forces driving DNA ejection, we performed *in vitro* measurements of DNA ejection in bulk and at the single-phage level. We present measurements on the dynamics of ejection and on the self-repulsion force driving ejection. We examine the role of ion concentration and identity in both measurements, and show that the charge of counter-ions is an important control parameter. These measurements show that the frictional force acting on the ejecting DNA is subtly dependent on ionic concentrations for a given amount of DNA in the capsid. We also present evidence that phage DNA forms loops during ejection; we confirm that this effect occurs using optical tweezers. We speculate this facilitates circularization of the genome in the cytoplasm.

**Keywords: bacteriophage, DNA ejection,** looping, optical tweezers, stall force

## Introduction

Bacteriophages have played a key role in the emergence of both molecular and physical biology. They were an essential component of the Hershey-Chase experiment (*1*), which established that DNA was the molecule of inheritance, and have since provided an important technology for cloning and protein expression (*2*). Phages have also been important model organisms for the study of macromolecular self-assembly; recent work has demonstrated assembly of viruses from their constituent components *in vitro* (*3*). Bacteriophages have also been instrumental in the field of systems biology, as evidenced by the dissection of the transcriptional program responsible for the decision between lysis and lysogeny (*5-9*). Finally, bacteriophages have played an important role in the field of single-molecule biophysics, as evidenced by the single-phage packaging experiments which revealed the large (many pico-Newton) forces generated by motors that package phage DNA into protein capsids (*4*). The study of the bacteriophage life cycle has yielded much insight in the physical biology setting and we anticipate that it will continue to serve as a useful model system.

Here, we highlight four recent experimental techniques that examine how phage DNA is packaged into and released from protein capsids. These four techniques include single-molecule packaging of DNA, single phage ejection measurements, osmotic suppression of DNA ejection, and cryo-electron tomography of partially packed phages. Single-molecule DNA packaging experiments have sought to elucidate the mechanics of how DNA gets packaged into the capsid; optical tweezers have been used to make systematic measurements of the packaging force for phage lambda and phi29 in a variety of different salt conditions (*10*). What these experiments demonstrated was that by increasing the charge of the salts, the genome is packaged at a higher velocity. Additionally, the internal force of the capsid DNA that resists packaging increases monotonically with a reduction in the charge of the salts.

These packaging experiments are complemented by single-phage ejection experiments, wherein phages are coerced into ejecting their DNA by addition of a trigger protein and then observed under a microscope with the aid of flow and DNA staining dyes (*11, 12*). These experiments illustrated the dynamics of the ejection process and have provided some insight into the frictional forces experienced by the DNA as it exits the capsid (*12, 13, 14*). The principal quantity measured in these experiments is the velocity of ejection as a function of the quantity of DNA inside the capsid. The observations garnered from these ejection experiments reveal (broadly speaking) a picture in which the forces responsible for the ejection process decay as a function of packaged length in the same way as the forces that build up during the packaging process depend upon the packaged length: a reduction in charge increases the speed of ejections. However, the available ejection data only reflect measurements on a few ionic conditions.

It is also possible to measure the force exerted by the DNA inside phage capsids during ejections by inducing phages to eject their DNA into a series of increasingly osmotically-resistive solutions (*15*). By varying the concentration of an osmolyte like polyethylene glycol (PEG), the resistive pressure is also varied. One can therefore measure the amount of DNA inside the phage capsid as a function of external pressure, and therefore how much force is generated by the DNA inside the capsid as a function of genome length (*16-18*). It has been proposed that the DNA self-repulsion force can help push the DNA into the bacterial cytoplasm (*19*), but a consensus has yet to be reached. Alternative models propose that highly osmotic environments serve as a condensing agent for DNA exiting phage and actually promote ejection (*18, 20*). Others posit that the osmotic difference (*21*) between the intra- and extracellular environment is sufficient to “flush” DNA from the phage into the host (*22*). However, when certain experimental conditions are met, by increasing the external pressure of the solution, phage ejections are inhibited (*20*). To our knowledge, a systematic study of how different salt conditions influence ejections at different resistive pressures has yet to be undertaken.

Another way to study ejection and packaging is by looking at the structure of the DNA itself inside the capsid. Cryo-electron tomography experiments in which the packaging process of phage phi29 was paused by freezing have provided information on the structure of viral DNA as a function of length of DNA inside the capsid (*23*). Others have investigated T5 (*24*), lambda (*25*, *26*), T3 (*27*), and P22 (see (*28*) and references therein). Recently, the interaxial spacing of condensed DNA was measured as a function of salts in order to corroborate viral DNA packaging theory (*29*).

The four experimental techniques described above have shed light on the forces, dynamics, and structure of DNA entering and leaving the viral capsid, and have been complemented by a vigorous modeling effort. These include simple structural models taking into account bending energies and *in vitro* measurements of electrostatic repulsion (*30-34*), density functional theory with few structural constraints that takes into account electrostatic repulsion/correlations and van der Waals forces (*35*) and molecular mechanics (*23*, *36, 37*). This body of work deepens our physical understanding of the packaging and ejection process.

In this work, we focus on the measurement of velocity and force of DNA ejection in bacteriophage lambda as a function of charge at fixed ionic strength. In previous experiments, it was shown that ejections are slower in the presence of magnesium as opposed to sodium (*13*); however, these experiments were not performed at fixed ionic strength, making it difficult to discern the role of other ion specific factors, such as charge. Our results demonstrate that the ionic composition controls the observed velocity and force, rather than ionic strength. We thus propose that the charge of the counter-ion species is an important control parameter for this system.

We also examined "looped" ejections, a type of ejection that was not analyzed in previous experiments and was assumed to be an artifact (*13*). In these ejections, the first portion of the DNA exiting the capsid is stuck at the site of origin, leading to a very bright piece of DNA. This attachment is eventually broken by the flow, fully unfolding the DNA. We demonstrate that looped ejections exhibit the same dynamics as unlooped ejections and that they are another ejection class, and not an artifact. We confirm this using optical traps. Finally, we provide technical improvements on the single-phage ejection assay, in particular showing that low dye concentrations are necessary to avoid spurious (non-LamB induced) ejections, as well as a reliable protocol for using field inversion gel electrophoresis to measure the amount of DNA ejected in bulk osmotic suppression assays, instead of spectrophotometric characterizations.

## Materials and Methods

*Phage Purification.* Lambda phage strain λcI60 was purified according to Sambrook (*38*). Briefly, stock phages were amplified by one round of plate lysis and one round of liquid phase lysis, followed by precipitation in PEG. Samples were then purified on a CsCl gradient followed by isopycnic centrifugation. Phages were dialyzed into the appropriate salt condition before use.

*LamB purification.* The membrane protein LamB was purified using a modified version of a protocol of Keller *et al* (*39*). The *E. coli* strain pop154 was grown overnight in LB + 0.2% maltose at 37 $^{o}$C. The cells were pelleted at 3000 g at 4 $^{o}$C for 30 minutes and re-suspended in 50 mM sodium phosphate pH 7.5, 100 mM NaCl, 2 mM EDTA, 5% sucrose. The re-suspended cells were subsequently lysed in a French press.

The lysate was again centrifuged at 5000 g at room temperature for 10 minutes. The supernatant was collected and the outer membrane fraction was pelleted by ultracentrifugation at 30000 rpm in a Beckman Ti-60 rotor, 18 $^{o}$C for 40 minutes. The pellet containing the membranes was resuspended in 20 mM sodium phosphate, pH 7.5, and 0.5% octylpolyoxyethylene (oPOE) in order to extract non-specific membrane proteins. The solution was incubated in a heat bath at 40 $^{o}$C for 50 minutes and again pelleted by ultracentrifugation. LamB was extracted by re-suspending the pellet in 20 mM sodium phosphate, pH 7.5, and 3% oPOE, incubating at 37 $^{o}$C for 40 minutes, and pelleting the resulting mixture by ultracentrifugation. The supernatant was dialyzed to 20 mM sodium phosphate, pH 7.5, and 1% oPOE and then loaded onto a GE MBPTrap column and eluted with 20 mM sodium phosphate, pH 7.5, 1% oPOE, and 20 mM maltose after washing twice with the same solution without maltose. The LamB solution was then dialyzed with 50 mM Tris, pH 7.5, 10 mM $MgSO_4$ (TM) and 1% oPOE. The resulting solution was diluted to 2 mg/mL. The concentration was assayed by measuring the absorbance at 280 nm and the purity was verified by SDS-PAGE.

*Single phage ejection assay.* See Fig. 1 A. We follow a protocol developed by Mangenot *et al.* (*12,13).* Microscope coverslips were cleaned by sonication in 1M KOH for 10 minutes followed by sonication in water for 10 minutes and dried on a hot plate. Glass slides were drilled using a diamond covered drill bit and 5 inches of tubing was attached to the glass slide using epoxy. The flow chamber was assembled using laser cut double-sided adhesive tape (*41*). A solution of $10^8$ - $10^{11}$ pfu/ml lambda phage and 20 nm fluorescent polystyrene beads (focusing aids) was incubated in the assembled flow chambers for 10 minutes. Once focused, the chamber was washed with 200 μL of buffer + 1% oPOE. Buffer consisted of either: 10 mM Tris, pH 7.5, 0 mM NaCl, 2.5 mM MgSO4; 10 mM Tris, pH 7.5, 2.5 mM NaCl, 1.875 mM MgSO4; 10 mM Tris, pH 7.5, 5.0 mM NaCl, 1.25 mM MgSO4; 10 mM Tris, pH 7.5, 7.5 mM NaCl, 0.625 mM MgSO4; or 10 mM Tris, pH 7.5, 10 mM NaCl, 0 mM MgSO4. The solution to induce ejection consisted of buffer, 1% oPOE, 1% glucose oxidase/catalase, $10^6$ diluted SYBR gold, and 1% LamB. Occasionally, 20 μg/mL acetylated bovine serum albumin, 3 mg/mL casein, and 80 μg/mL heparin were added to the buffer to block the glass surface to check if looping ejections disappeared. Calibration of lengths was performed as in (*13*).

*Microscopy.* Samples were imaged at 60X and 100X using a Nikon perfect focus system and either a SYBR gold or FITC filter set. Images were acquired with a Photometrics CoolSnap ES2 camera; the exposure time was 75 ms and the frame rate was ~4 fps.

*Image analysis.* A custom segmentation algorithm was developed using MATLAB. Ejection events were identified in each movie by local adaptive thresholding on the fluorescence intensity. DNA was identified in each ejection by using background subtraction, a Wiener filter for de-noising, and a custom Canny filter for edge finding. The long axis length, total intensity, and background (the mean intensity of non-DNA related pixels) were extracted for each ejection. Trajectories were subjected to a 3 bin median filter for de-noising.

*Force dependence assay*. We follow a protocol adapted from Evilevitch *et al* (*15*). Phages were quickly swirled in a solution of 1% oPOE, 1% LamB, 1% DNAse and appropriate buffer containing PEG at 37 $^{o}$C and incubated for 1 hour at 37 $^{o}$C (*17*). The solution was then incubated in a 65 $^{o}$C water bath for 50 minutes to break open capsids and inactivate DNAse. DNA in capsids was then extracted by a 1:1 phenol:chloroform extraction followed by chloroform extraction and finally ethanol precipitation. DNA lengths were assayed using field inversion agarose gel electrophoresis (100 V forward 0.8 sec, 60 volts backwards 0.8 sec, for 7-10 hours). After staining with ethidum bromide, migration distance was quantified using MATLAB.

*Optical trapping.* See Fig. 1 B. Phages were biotinylated after a standard plate lysis by incubation with 10 μM of sulfo-LC-biotin, after first dialyzing for 48 hours with 100 mM sodium bicarbonate, pH 8.2. The phages were then purified on a CsCl gradient. Biotinylated phages were incubated with 2 μm streptavidin coated polystyrene spheres for a few hours at room temperature on a rotating rack. The streptavidin spheres were washed by resuspension in TM 3 times and diluted 10-fold into TM during use. The phage ejection solution (see *single phage ejection assay* above) was injected into one port, and the phages conjugated to streptavidin spheres were injected into the other port at 10 μL/min. A 1064 nm laser at 100 mW was used to trap phage-bound spheres, which were moved across the boundary layer of the two solutions. The microscope was home built around a 60X IR corrected water immersion objective adapted with a 200 mm focal length tube lens, and imaged with an Andor Ixon EMCCD camera. The schematic of the chamber in Fig. 1 B, was adapted from (*41*).

**Results and Discussion**

Our first set of experiments focused on optimizing the single phage ejection assay with regards to SYBR gold concentration, since we noticed that phages spontaneously eject their DNA in high concentrations of SYBR gold. SYBR gold is thought to affect the mechanical properties of DNA since other DNA stains have been shown to affect DNA's persistence length (*42*). To investigate the origin of these ejections and to test the hypothesis that they are triggered by SYBR gold, we incubated phages in different salts and different amounts of SYBR gold and subsequently measured the number of active phages by titering. The results are given in Table 1; errors in the titer values follow $\sqrt{N}$ (counting) statistics. For high amounts of SYBR gold (1:10000), we see a steep drop in the number of active phages, between 2- and 10-fold for the buffers containing magnesium. The drop is particularly drastic in pure sodium (1000-fold), a condition thought to increase the DNA pressure inside the capsid. One explanation is that sodium in concert with SYBR gold increases the DNA pressure enough to destabilize the phage. Alternatively, magnesium and SYBR gold could compete with each other for binding sites on the DNA, providing a protective effect. Another possibility is that SYBR gold acts by a mechanism similar to LamB. Regardless, the drop in titer is consistent with the hypothesis that SYBR gold induces ejections. When the SYBR gold concentration is reduced (1:1000000), we recover almost the entirety of the original titer value. We concluded that it was preferable to perform the experiment at this lower concentration to minimize the effect of SYBR gold. The concomitant reduction in signal did not affect our DNA segmentation algorithms.

When performing the single phage ejection assay (Fig. 1 A), we always noticed two types of ejections - ejections that look similar to those previously reported and those that revealed a different type of dynamics that have the appearance of a "partial ejection" and were previously viewed as artifactual (*13*) (Fig. 3 A). Furthermore, we always saw spurious ejections of both types. In our investigations into the origins of the spurious ejections, we found that, under certain salt conditions (Table 1), we saw more "artifact" ejections after addition of LamB (up to 3 fold increase) in the single phage ejection assay. This led us to question their designation as "artifact."

However, it is well known that DNA will stick to glass beads (*43*), and the previous experiment does not rule out DNA sticking to the microscope coverslip, although this seems unlikely since the addition of LamB (in excess) would tend to block sites on the glass for the ejecting DNA to bind to. The stickiness of glass was confirmed by incubating lambda phage DNA on freshly cleaned coverslips; upon observation, the glass was covered in DNA (data not shown). Thus as another test, we performed the ejection assay in the presence of the surface blocking agents BSA, casein, and heparin. These blocking agents significantly reduced DNA's affinity to the surface (*44*) – when bare lambda DNA was again incubated in a flow chamber with these blocking agents, only a single strand was found after futile searching efforts (data not shown). However, even with these blocking agents, "artifactual" ejections were seen in the phage ejection assay. Fortunately, the content of the "artifact" ejections could be deduced from serendipitous ejections: occasionally, it could be seen that the DNA was in fact in a looped state. This was quantified by an intensity histogram and is shown in Fig. 2 A. We therefore term this class "loop" ejections in contrast to the ejections previously characterized, which we call "continuous" ejections. The addition of LamB tends to increase the amount of looping ejections (Table 1) which suggests that looping ejections are physiological.

We further demonstrated that the looping ejections were not surface mediated by ejecting biotinylated phages off of optically trapped streptavidin coated polystyrene spheres (Movie S1). A dual-syringe pump pushes fluid into the chamber (Fig. 1 B). Using an optical trap, we grabbed hold of a microsphere, and moved it across the laminar flow boundary layer into the ejection buffer. The continuous flow system limits diffusive mixing at the meeting juncture. After a LamB trimer finds its target, a phage ejects and we can monitor the ejection (Fig. 2 B). We found that the looped ejections were as common as continuous ejections. These experiments support the hypothesis that loops occur when the exiting DNA sticks to the LamB or the phage capsid.

We next characterized the dynamics of both the looped and continuous ejections under different ionic compositions, again using low concentrations of SYBR gold as described above (Fig. 1 A). We first flowed in the dye front to visualize any spontaneous or SYBR gold-induced ejections. Even at a 10 fold lower concentration of SYBR than previously reported (*13*), there were still some spurious ejections, although much less than that would occur with SYBR gold concentrations used previously (*13*). Any lower concentration of dye would have precluded accurate segmentation of the ejecting DNA. After a period of time upon which the rate of spurious ejections was observed to be steady, we added LamB to the solution. A sudden large increase in the rate of visualized ejections indicated that LamB was inducing a large fraction of subsequent ejections. However, it should be noted that the dynamics of ejections induced by SYBR gold and those induced by LamB are indistinguishable. The experiment was repeated for 5 different salt conditions, in which we systematically varied the amount of sodium and magnesium ions while keeping the ionic strength constant.

We analyzed as many ejections within a field of view as possible – reasons for excluding an ejection from analysis included overlapping with another ejecting phage, photo-damage during an ejection, or a looped ejection that did not completely unfold. Ejections extracted from the field of view were partitioned into two categories – "continuous" and "looped". These are highlighted in Fig. 3, A and B, respectively. To quantify the dynamics, we designed a custom segmentation algorithm for the DNA strands. The relation between length of DNA in basepairs and in pixels was determined by measuring, under flow, different known lengths of lambda DNA cut with restriction enzymes, and fitting the result to a quadratic function (Fig. S1 in the Supporting Materials). This calibration disentangles polymer stretch and shear flow from actual DNA length (*45*). The DNA's long axis length for each time point in a trajectory was then interpolated according to the quadratic function and normalized by 48.5 kbp, the full length of lambda phage DNA. This calibration is only accurate for continuous ejections; for looped ejections it is necessary to use intensity to discern the amount of DNA ejected. To do so, we take the total intensity of the DNA above background and normalize by the maximum intensity observed in a trajectory. This was done for both continuous and looped ejections (Fig. 3, C and D, Fig. S2, A-E).

A representative example of the trajectories is shown in Fig. 3, C and D – trajectories for each ionic condition are in Fig. S2, A-E. The fraction of DNA ejected is plotted versus time – the green curve uses fluorescence intensity to quantify total DNA ejected while the blue curve uses the calibrated length. It is clear in "looped trajectories" that the DNA is fact pinned at one end, as the trajectories exhibit a discontinuity in length at half-maximal ejection. A comparison between using calibrated DNA lengths and intensity as a measure of total ejected DNA is shown in Fig. S3, A-E. Intensity measures give a larger estimate for the velocity of DNA ejected at early times because our segmentation algorithm misses the first bits of DNA ejection, and so changes in DNA length appear relatively large at small lengths; whereas at late stages in the ejection, length and intensity measures are equivalent metrics for amount of DNA ejected.

For each condition, we binned the trajectories to compute the average velocity as a function of DNA remaining in the capsid. This is plotted in Fig. 3, E and F; Fig. 3 E was computed using calibrated length while Fig. 3 F was computed using intensities. Velocity plots for individual salt conditions are in Fig. S4, A-E. From our analysis of the dynamics, we observed that individual trajectories were very similar to each other. As with previous work (*13*), this leads us to conclude that we are observing the same intrinsic dynamical process and that differences between trajectories are likely the result of measurement error. Secondly, our systematic perturbation of the sodium and magnesium concentrations - at fixed ionic strength - showed a consistent trend. As the amount of magnesium was lowered and the amount of sodium was increased, the speed of the ejections increased (Fig. 3, E and F). From this, we conclude that it is the type of ion, not simply how much is present, that is an important control parameter that governs DNA ejection dynamics. We posit that the role of the positive ions is to screen the negative charge of the DNA backbone, which suggests that the charge of the counter-ions is the property of interest.

Additionally, we observe that the continuous and looped ejections have velocity curves that are within error of each other (Fig. 3, E and F). This suggests that the ejection mechanism for these two classes of ejections are very similar, if not identical, and that what happens to the first segment of DNA that exits the capsid determines whether an ejection will be continuous or looped.

To continue the physical characterization of the ejection process, we performed osmotic suppression experiments to determine the forces driving ejection in each of the different salt conditions. Atmospheric pressure varies with PEG 8000 concentration as given in (*46*). Previous experiments have measured the amount of DNA ejected by pelleting phage capsids and measuring the absorbance at 260 nm. This measures the amount of DNA that has been chopped up by the DNAse, which is assumed to be entirely ejected DNA. While the mass of DNA is thus accurately measured, it is not clear if the mass of ejected DNA can be quantitatively related to the length of DNA remaining in the phage capsids (see Supplementary Information). We used an alternative approach – purifying the DNA inside the capsids by standard extraction techniques and measuring its length directly by using field inversion gel electrophoresis (Fig. S5) (*15*). This method is advantageous because we no longer need to assume that a constant fraction of phages eject for each PEG concentration. Care was taken to avoid centrifugation, as it tends to fragment DNA (data not shown). The pressure driving ejection is shown in Fig. 4 A. The general trend is that decreasing magnesium and increasing sodium increases the driving force, although the 5, 7.5, and 10 mM Na samples appear to be within error of each other. We observed that 40 atmospheres (14 pN) of pressure was not enough to stall ejections in some salt conditions, which is consistent with theoretical expectations (*13*). Atmospheres can be converted to force by multiplying the pressure by the end-on area of DNA, assuming a radius of 1 nm.

We can determine how the mobility, denoted by $\sigma(l)$, of DNA inside the capsid depends on the salts (Fig. 4 B) by assuming a linear relation between force and velocity, $F(l) = v(l)/\sigma(l)$, where $l$ is the DNA length in the capsid, $F$ is the force as measured in Fig. 4 A (multiplied by the area of end-on DNA), and $v$ is the velocity as measured in Fig. 3 E and F. As previously observed (*13*), the mobility depends strongly on the amount of DNA remaining in the capsid. We are unable to conclude that mobility is dependent on salt condition, although there is a slight trend that the measured mobility is higher for solutions with more multivalent salts.

**Conclusion**

We investigated the dynamics of DNA ejection in a variety of different salts, and also performed measurements on the self-repulsive force driving this process. We demonstrated that the counter-ion charge is an important control parameter for this class of experiments. These measurements should be of use to theorists working in physical virology. Of particular interest would be models that predict the force driving ejection in different salts, in addition to a better understanding of the time scale of the ejection process. We also investigated the origin of looped ejections. From our data, we can develop several possible models for these processes. It is possible that the end of the DNA gets stuck on the tail of the phage or LamB as it comes out, and that the rest of the genome subsequently comes out of the tail, accounting for the DNA loops occasionally seen (Fig. 2 A). This may make it possible for phage to circularize even faster as the genome rushes to get inside its host, *E. coli*, as it is theorized that lambda delivers its genome directly into the cytoplasm without going through the periplasmic space (*47*). Although the observation of looped ejections was first made in 2007 (*13*), this is the first attempt to explain them and propose that they might represent an important part of phage lambda's life cycle.

Of particular note is the relation of the current work to (*10*), which is an investigation of the packaging process as a function of salts; they note that by increasing the amount of multivalent cations in their packaging assay, the virus packages faster, and requires less force at an equivalent fraction of DNA packaged. This suggests that increasing magnesium reduces the amount of intrastrand DNA repulsion. In this current study, we see a similar effect; namely, an increased force is required to stall ejecting DNA at the same length, with increased monovalency. However, the mobility during DNA ejection appears to be independent of salt composition at our detection resolution. Notably, the speed of ejection is faster during ejection in the presence of larger fractions of monovalent cations, which is consistent with the packaging data. Altogether, this work is a step forward in understanding the forces at play when bacteriophage lambda infects bacteria.

**Acknowledgements**
The authors thank the members of the Physiology course at the Marine Biology Laboratory in Woods Hole, MA, Nicolas Chiaruttini, Paul Grayson, Zenan Chang, Alexander Grosberg, Ian Molineux, Michael Rubinstein, Virgile Viasnoff, William Gelbart, Charles Knobler, and members of the Phillips lab.

D.W. and D.V.V. were supported by the NIH Medical Scientist Training Program Fellowship. D.V.V. was supported by a Yaser Abu-Mostafa Hertz Fellowship and the NIH Director's Pioneer Award. RP, D. W. and D. V. V. were supported by NSF grant number 0758343 and the NIH Pioneer Award.

**Tables: Phage survival in SYBR gold**

| Ionic composition | Titer ($x 10^{11}$) | Titer in high SYBR ($x 10^{11}$) | Titer in low SYBR ($x 10^{11}$) | % looped in low SYBR with LamB | % looped in low SYBR with no LamB |
|---|---|---|---|---|---|
| 50 mM Tris<br>10 mM Mg<br>0 mM Na | 1 | 0.41 | 0.55 | 88 | 35 |
| 10 mM Tris<br>2.5 mM Mg<br>0 mM Na | 0.76 | 0.07 | 0.59 | 58.8 | 13.2 |
| 10 mM Tris<br>1.875 mM Mg<br>2.5 mM Na | 0.90 | 0.10 | 0.86 | 57.6 | 25 |
| 10 mM Tris<br>1.25 mM Mg<br>5 mM Na | 0.79 | 0.03 | 0.97 | 50 | 48 |
| 10 mM Tris<br>0.625 mM Mg<br>7.5 mM Na | 0.93 | 0.08 | 1.10 | 44.9 | 20 |
| 10 mM Tris<br>0 mM Mg<br>10 mM Na | 0.69 | 0.001 | 0.55 | 55.3 | 6.9 |

**Table 1.** SYBR gold triggers DNA ejection independent of LamB; the presence of LamB influences the character of ejection. (a) Columns 1-3: Titers of phage lambda solutions as a function of buffer composition and SYBR gold concentration. CsCl purified lambda phage were dialyzed in the appropriate buffer by washing and spin filtering. SYBR gold was added to bring the solution to the right concentration (none, high (1:10,000) and low (1:1,000,000)). The number of viable phage remaining was measured by titering. High concentrations of SYBR gold reduce the titer by more than an order of magnitude, especially in the absence of magnesium. This reduction in titer is because SYBR gold induces the bacteriophage to eject their DNA. Lowering the concentration of dye recovers the original titer levels. (b) Columns 4-5: Character of ejection in the presence and absence of LamB. A single phage ejection assay was run for each condition with and without LamB. The number of ejections, both continuous and looped, was counted manually and is reported above. The fraction of ejections that are looped is higher in the presence of LamB.

**Figure Legends:**

**Figure 1:** Schematics for observing DNA ejection by single phage lambda virions *in vitro*. DNA was visualized using fluorescence microscopy. (A) Ejection at a surface. A flow chamber was incubated with lambda phage and the viruses were allowed to non-specifically settle on to the surface. DNA ejection was triggered by flowing in LamB; the ejected DNA was visualized using SYBR gold. The flow field extends the ejected DNA, allowing quantification of the amount of DNA ejected by either measuring the length in pixels or measuring the total intensity. (B) Ejection on a bead. A parallel ejection assay using an optical trap was developed. Biotinylated phages were bound to streptavidin coated polystyrene beads. A dual-input flow chamber was used; beads coated with lambda phage entered the chamber through one input and an ejection buffer consisting of LamB and SYBR gold entered through the other. In the presence of flow, the two fluids are well separated. An optical trap was used to move a phage-coated bead into the ejection buffer to initiate DNA ejection.

**Figure 2:** Looping during DNA ejection. (A) A montage of loops observed during the single phage ejection assay. An intensity histogram constructed from one of the examples demonstrates the presence of loops during *in vitro* ejections. (B) Montage observation of a looped ejection during an optical trapping experiment. Phages were ejected off of optically trapped beads to suppress effects caused by the surface. Looped ejections are still observed, suggesting that they are not the byproduct of surface effects. The bead is 2 microns in diameter.

**Figure 3:** Trajectories of single phage ejections and the ion-concentration dependence of the ejection dynamics. (A) Montage of continuous and (B) looped ejections. Movies of individual phages ejecting their DNA. The identified pieces of DNA are outlined in red. (C, D) Sample trajectories for continuous and looped ejections. The amount of DNA ejected was quantified by measuring either the calibrated long axis length or the total intensity above background of the segmented region. Lengths were determined by calibrating against restriction enzyme digested lambda DNA at known lengths, as described in the text. Individual trajectories are normalized by the maximum length/intensity observed during an individual trajectory to obtain the fraction ejected. Two qualitatively different ejections are observed – continuous ejections and looped ejections. The looped ejections reach half-maximal length before unfolding. The total intensity provides a way to quantify the total amount of DNA outside the capsid. (E) Velocity of ejection as a function of buffer composition for continuous trajectories and (F) looped. For each trajectory, the velocity of the DNA (measured using total intensity) at different landmark lengths was recorded. The mean velocity and the standard error of the DNA at each landmark length are plotted for each buffer condition. There is a clear trend in the data – buffers with more magnesium ions have a longer timescale for DNA ejection. Continuous and looped ejections have similar ejection dynamics in all buffer conditions.

**Figure 4:** The force driving DNA ejection. Osmotic suppression experiments were performed in different buffer conditions to measure the force driving ejection. The fraction of DNA ejected was determined by measuring the amount of capsid DNA remaining using field inversion gel electrophoresis. (A) Increasing monovalent ion concentration raises the stall force of ejection at each length of DNA. (B) The mobility is measured as $v(l)/F(l)$. The trend is less clear that regarding mobility and ion composition, though it appears that higher amounts of divalent cations increase mobility.

1.

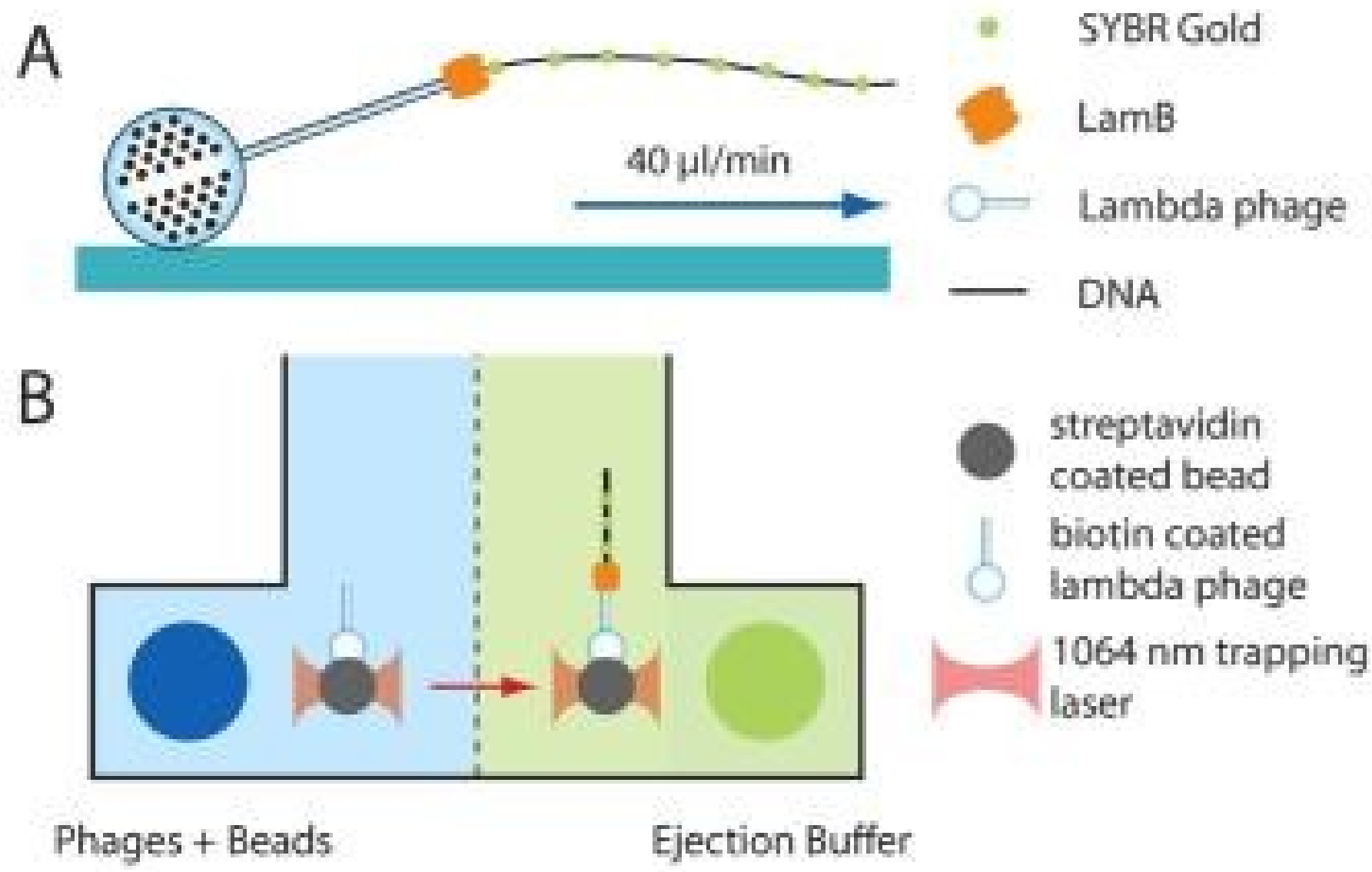
A
SYBR Gold
LamB
40 µl/min
Lambda phage
DNA
B
streptavidin coated bead
biotin coated lambda phage
1064 nm trapping laser
Phages + Beads
Ejection Buffer

2.

A

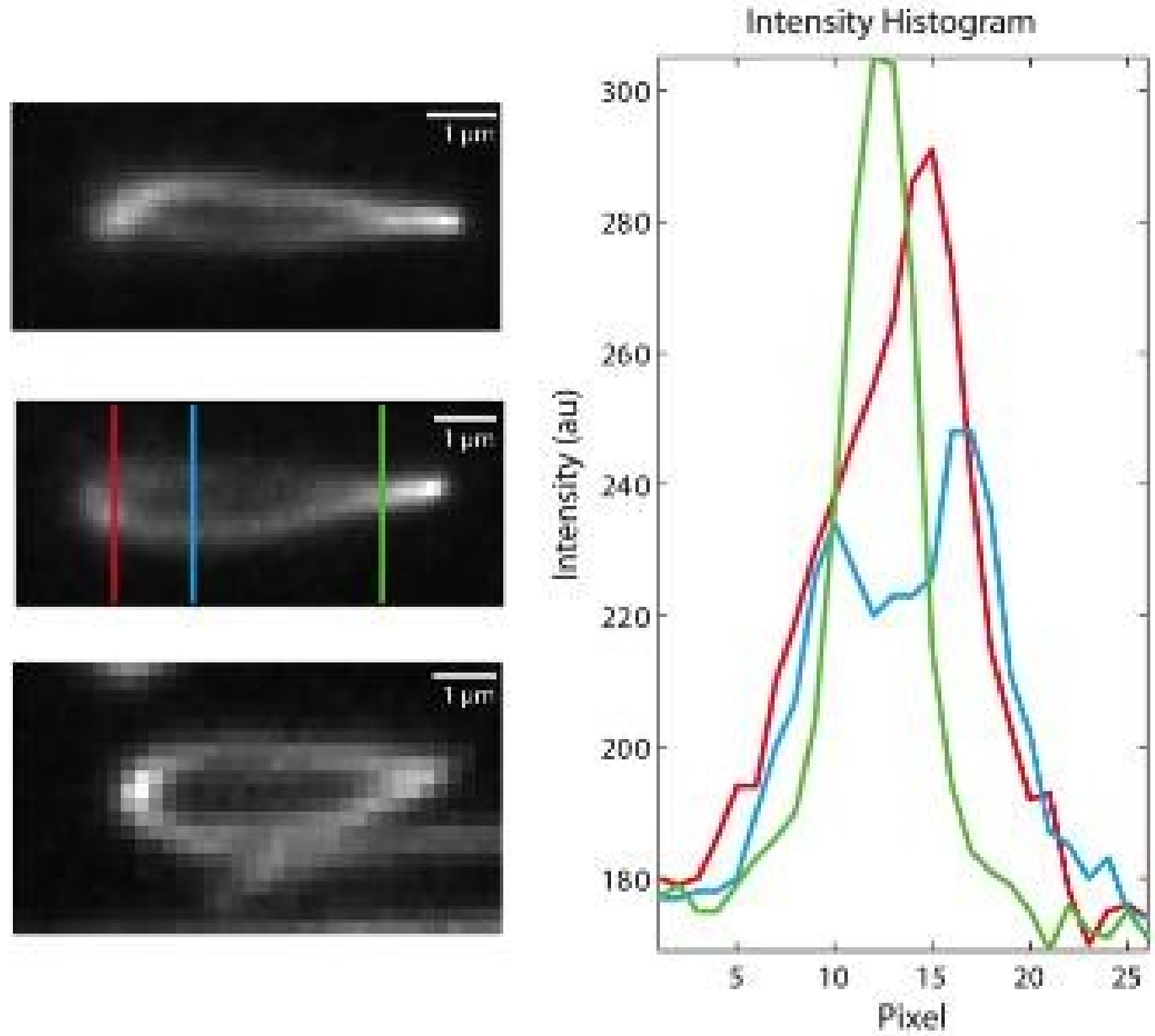

1 µm
1 µm
1 µm
Intensity Histogram
Intensity (au)
Pixel

B

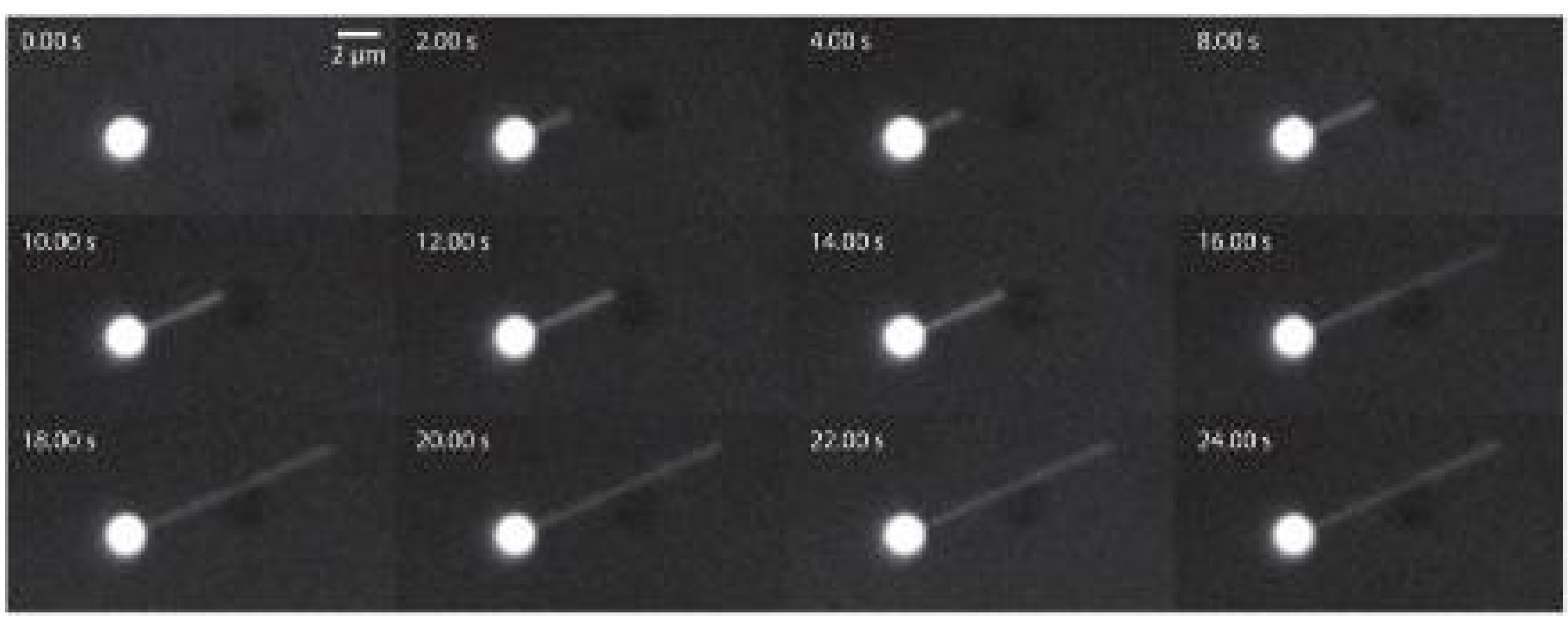

0.00 s
2 µm
2.00 s
4.00 s
8.00 s
10.00 s
12.00 s
14.00 s
16.00 s
18.00 s
20.00 s
22.00 s
24.00 s

3.

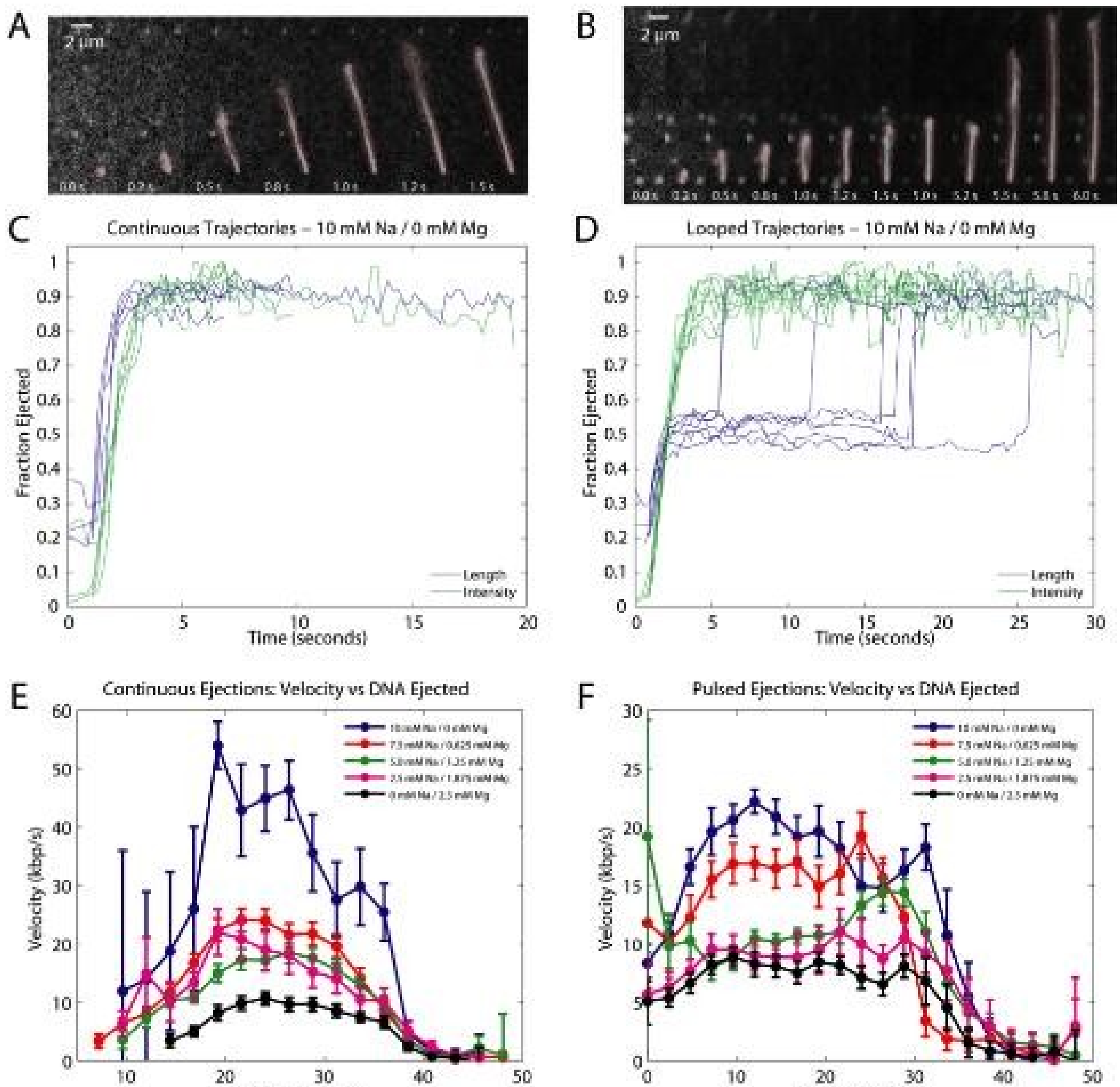

A
2 μm
0.0 s
0.2 s
0.5 s
0.8 s
1.0 s
1.2 s
1.5 s
B
2 μm
C
Continuous Trajectories – 10 mM Na / 0 mM Mg
Fraction Ejected
Time (seconds)
Length
Intensity
D
Looped Trajectories – 10 mM Na / 0 mM Mg
Fraction Ejected
Time (seconds)
Length
Intensity
E
Continuous Ejections: Velocity vs DNA Ejected
Velocity (kbp/s)
DNA Ejected (kbp)
F
Pulsed Ejections: Velocity vs DNA Ejected
Velocity (kbp/s)
DNA Ejected (kbp)

4.

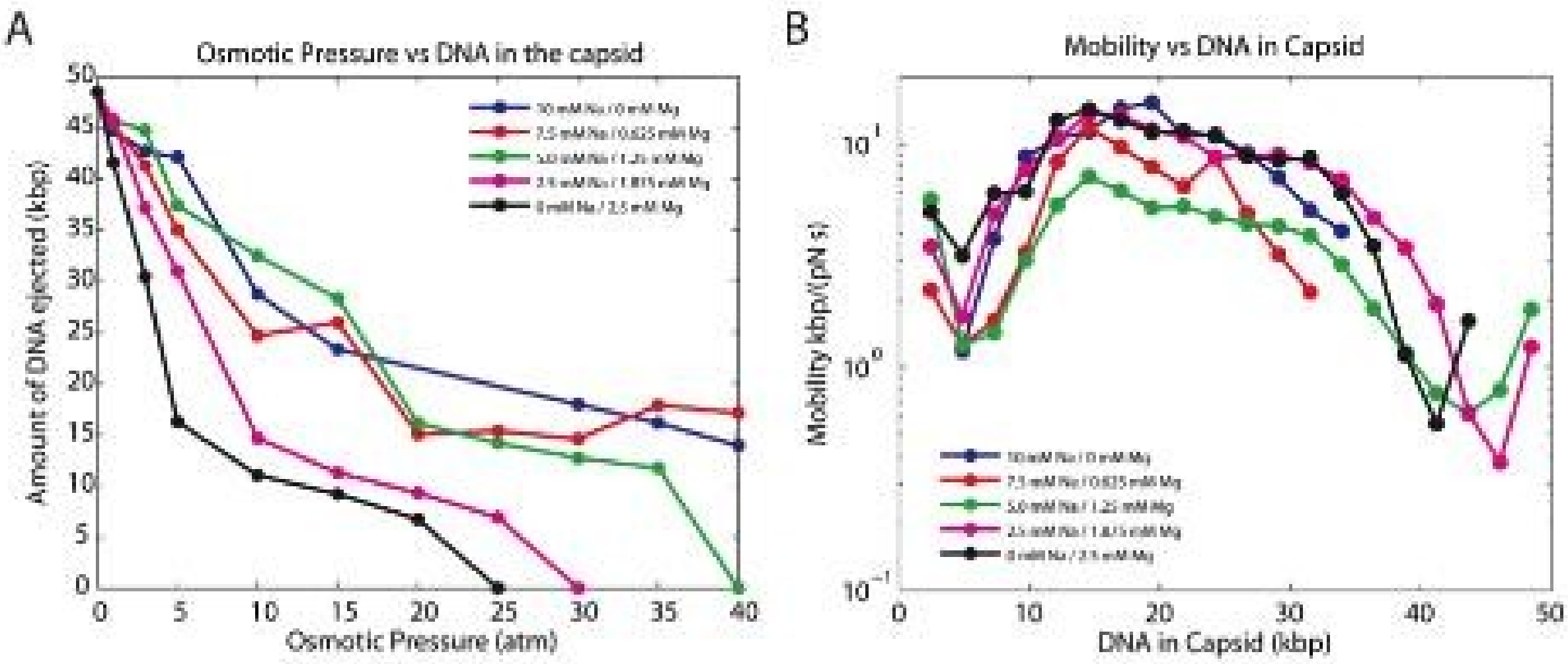
A
Osmotic Pressure vs DNA in the capsid
Amount of DNA ejected (kbp)
Osmotic Pressure (atm)
10 mM Na / 0 mM Mg
7.5 mM Na / 0.625 mM Mg
5.0 mM Na / 1.25 mM Mg
2.5 mM Na / 1.875 mM Mg
0 mM Na / 2.5 mM Mg
B
Mobility vs DNA in Capsid
Mobility kbp/(pN s)
DNA in Capsid (kbp)
10 mM Na / 0 mM Mg
7.5 mM Na / 0.625 mM Mg
5.0 mM Na / 1.25 mM Mg
2.5 mM Na / 1.875 mM Mg
0 mM Na / 2.5 mM Mg

**Supplemental Information**

Fig. S1 is the calibration of measured length of DNA in the microscope to actual length of DNA in kbp as determined by restriction enzymes (Grayson *et al*., 2007). The result is that length measured in the microscope is roughly quadratic with actual length in base pairs. We use this calibration to determine the actual length of ejected DNA in Fig. 3, C and E, and Fig. S2.

Herein we plot all trajectories (Fig. S2) used to calculate velocities in Fig. 3, E and F. In Fig. S3 and S4 we describe how the two measures of DNA ejected, length and intensity, are related to each other. In brief, fluorescence intensity consistently overestimates DNA at small lengths, and underestimates at long lengths (see Fig. S3 legend for details). Fig. S2 demonstrates characteristic “continuous” and “looped” trajectories for different salt conditions. Looped trajectories clearly show a discontinuous change in length as a function of time, while the intensity measurements suggest a continuous exit of DNA from the capsid as a function of time and that all of the DNA has exited the capsid even though the length is much shorter than the fully stretched DNA. This suggests that for the “looped” trajectories the DNA is pinned to the capsid (Fig. 2, B) and ultimately unfolds due to the presence of the flow, resulting in a discontinuous change in length as shown in the figure.

Evilevitch *et al*., 2003, developed a method to measure the amount of DNA remaining in phage capsids after incubation with LamB in the presence of osmolytes using spectophotometry (absorbance at 260 nm) and by field inversion gel electrophoresis (Evilevitch *et al*., 2005); in the former measurement, the fraction of phages that did not eject was assumed to be constant at different concentrations of PEG. On the basis of this assumption, the $A_{260}$ measurement can be directly converted from mass to length; however, if the fraction of ejected phages changes from preparation to preparation, spectophotometry will not produce an accurate measure of length. Moreover, sample to sample differences in preparation (DNA precipitation yield, especially) make it difficult to conclude that there is a constant unejected fraction – this is shown in Figure 5 of Evilevitch *et al*., 2005, where there is up to a 30% variability in $A_{260}$ measurements. To circumvent this issue, we chose to visualize the DNA length directly by performing agarose gel electrophoresis (Figure S5) only. We see clearly that increasing the external pressure inhibits DNA ejection. By measuring the migration distance in each lane, we can relate the pressure exerted on the phage to the amount of DNA that was not remaining in the capsid, and in this fashion derive a pressure corresponding to the DNA left in the capsid as a function of amount of DNA ejected (Fig. 4, A).

Above each lane in Fig. S5 is indicated the amount of external pressure induced by the presence of the PEG as measured in atmospheres. The relation between PEG concentration and pressure is described in Michel, 1983. Lanes labeled “L” correspond to the DNA ladder, with the marker lengths noted in (Fig. S5, A). Increasing amounts of external pressure results in increasing amounts of DNA left inside the capsid after ejection. Migration distance in each lane was measured by looking for the maximum fluorescence peak relative to the 48.5 kbp peak, and comparing that to the corresponding ladder. Since electrophoresis was performed in many different batches, we present the results in such a way that the relevant pressure experiment in the appropriate salt condition is to the right of its corresponding ladder. Electrophoresis parameters: 100 V forward, 0.8 sec; 60 V backwards, 0.8 sec; 7-10 hours.

Using the results from Figs. S3 and S4 (continuous velocities only), the end-on area of a double-stranded piece of DNA (radius = 1 nm), and assuming a linear relation between force and velocity, we arrive at the amount of intra-capsid friction that the DNA encounters at different lengths (Fig. 4B).

[Movie available upon request].

Movie S1. Biotinylated lambda phages ejected off of a 2 micron sphere, held in an optical trap. The schematic for the experiment is shown in Fig. 1, B. Both continuous and looped ejection events are evident.

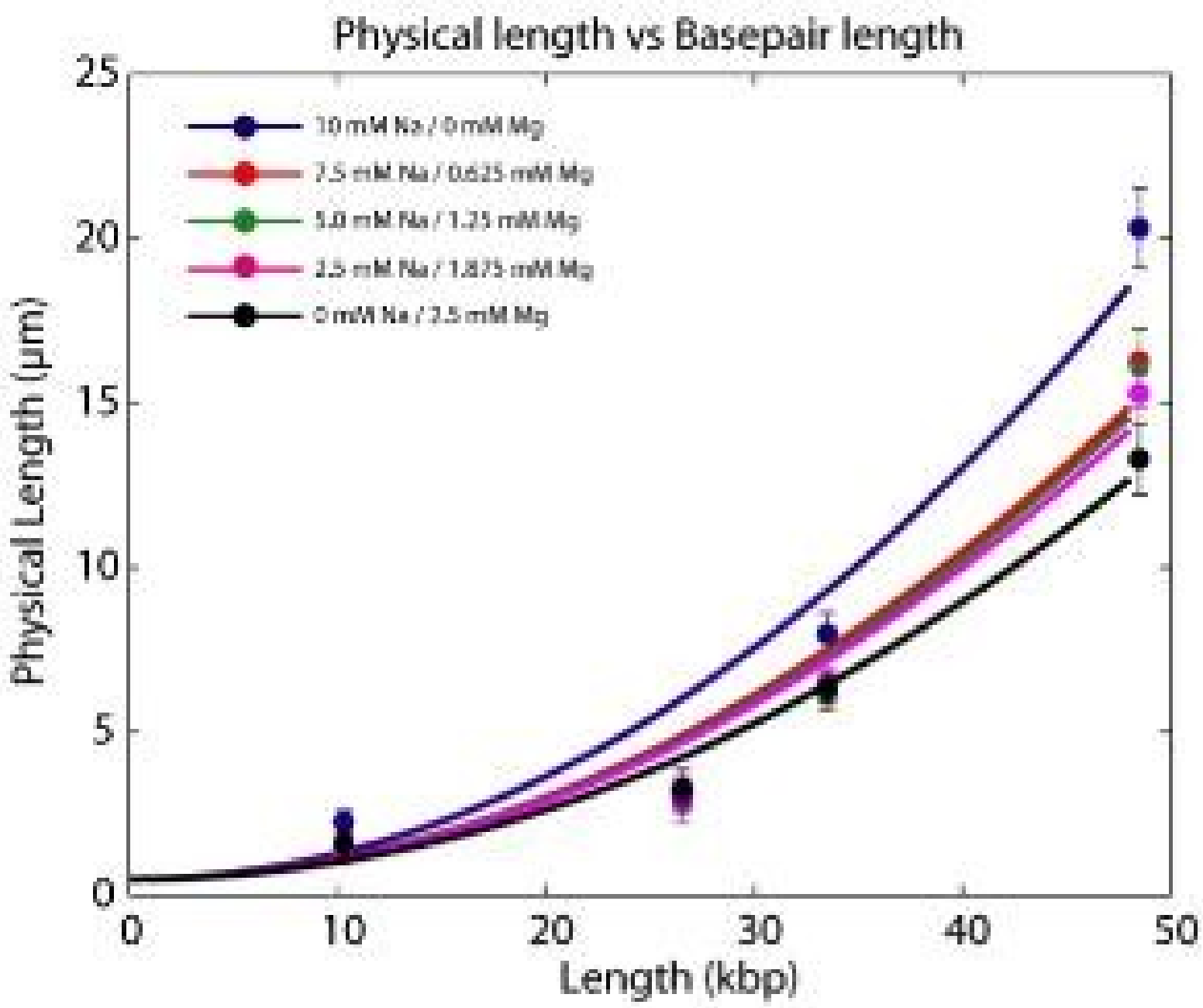

Figure S1. Calibration of tethered DNA length. To convert the physical length of DNA to the length in kbp, we performed a calibration as done by Grayson *et al.,* 2007. Briefly, biotinylated lambda genomes were either treated with restriction enzymes (XhoI, NsiI, and PmlI) or kept untreated. The DNA was attached to a glass coverslip through a biotin-streptavidin bond. The DNA was extended with a 40 µl/min flow and the lengths of different pieces of DNA were measured manually using ImageJ. The mean length was fit to a quadratic function; error bars represent the standard deviation of the measured length.

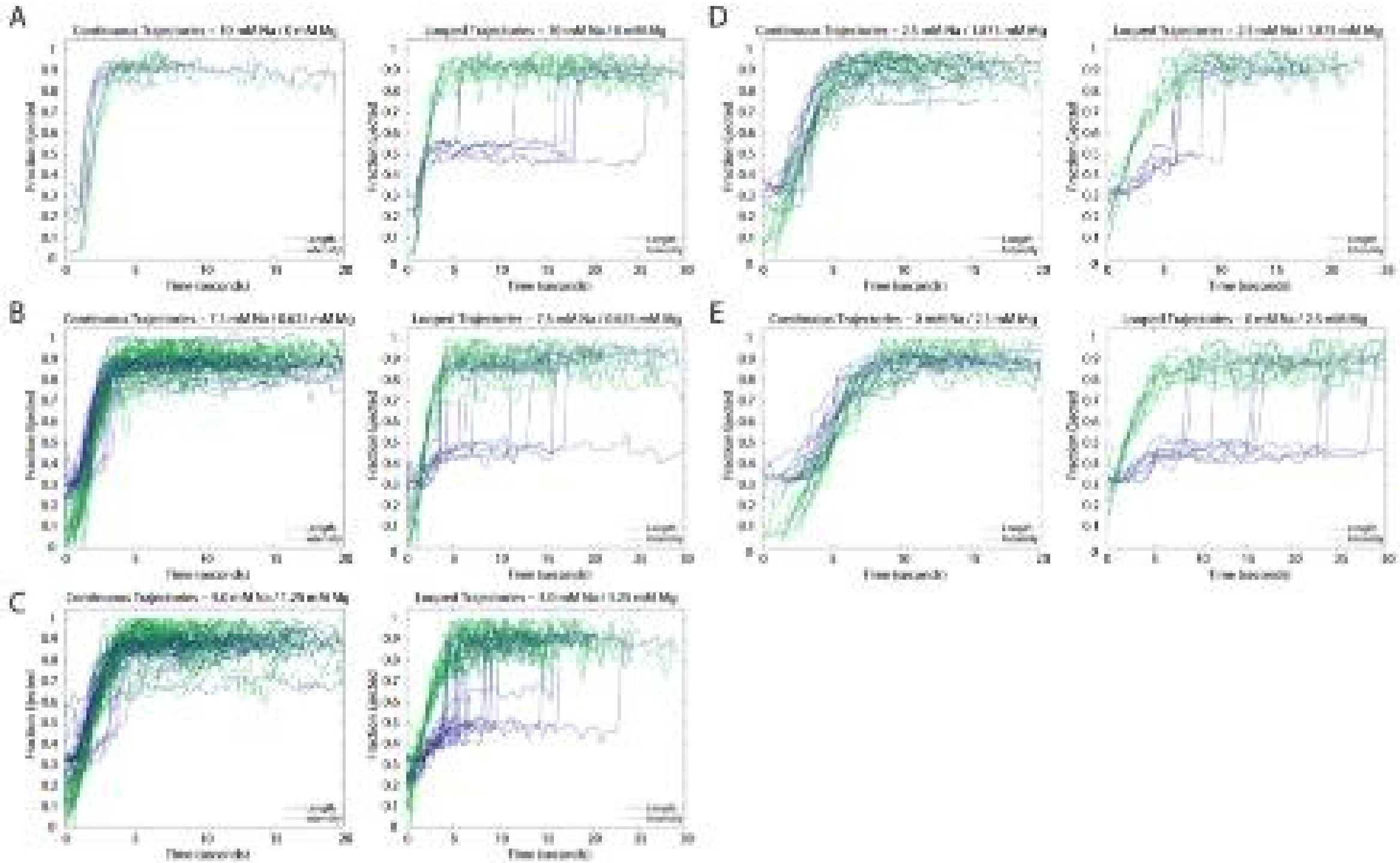

Figure S2. Trajectories used in the paper for all salt conditions. Green: intensity, blue: length. Intensity is the summed intensity of the DNA labeled by SYBR gold. The length is calibrated by measuring defined-length lambda DNA, via restriction digests, under the same flow conditions. Length is then calculated by interpolation. Looping trajectories uniformly unfold at half the maximum length. (A) Salt condition: 10 mM Tris, pH 7.4, 10 mM NaCl, 0 mM MgSO4. Number of trajectories: 5 continuous and 8 looped. (B) Salt condition: 10 mM Tris, pH 7.4, 7.5 mM NaCl, 0.625 mM MgSO4. Number of trajectories: 36 continuous and 13 looped. (C) Salt condition: 10 mM Tris, pH 7.4, 5.0 mM NaCl, 1.25 mM MgSO4. Number of trajectories: 43 continuous and 25 looped. (D) Salt condition: 10 mM Tris, pH 7.4, 2.5 mM NaCl, 1.875 mM MgSO4. Number of trajectories: 14 continuous and 6 looped. (E) Salt condition: 10 mM Tris, pH 7.4, 0 mM NaCl, 2.5 mM MgSO4. Number of trajectories: 15 continuous and 10 looped.

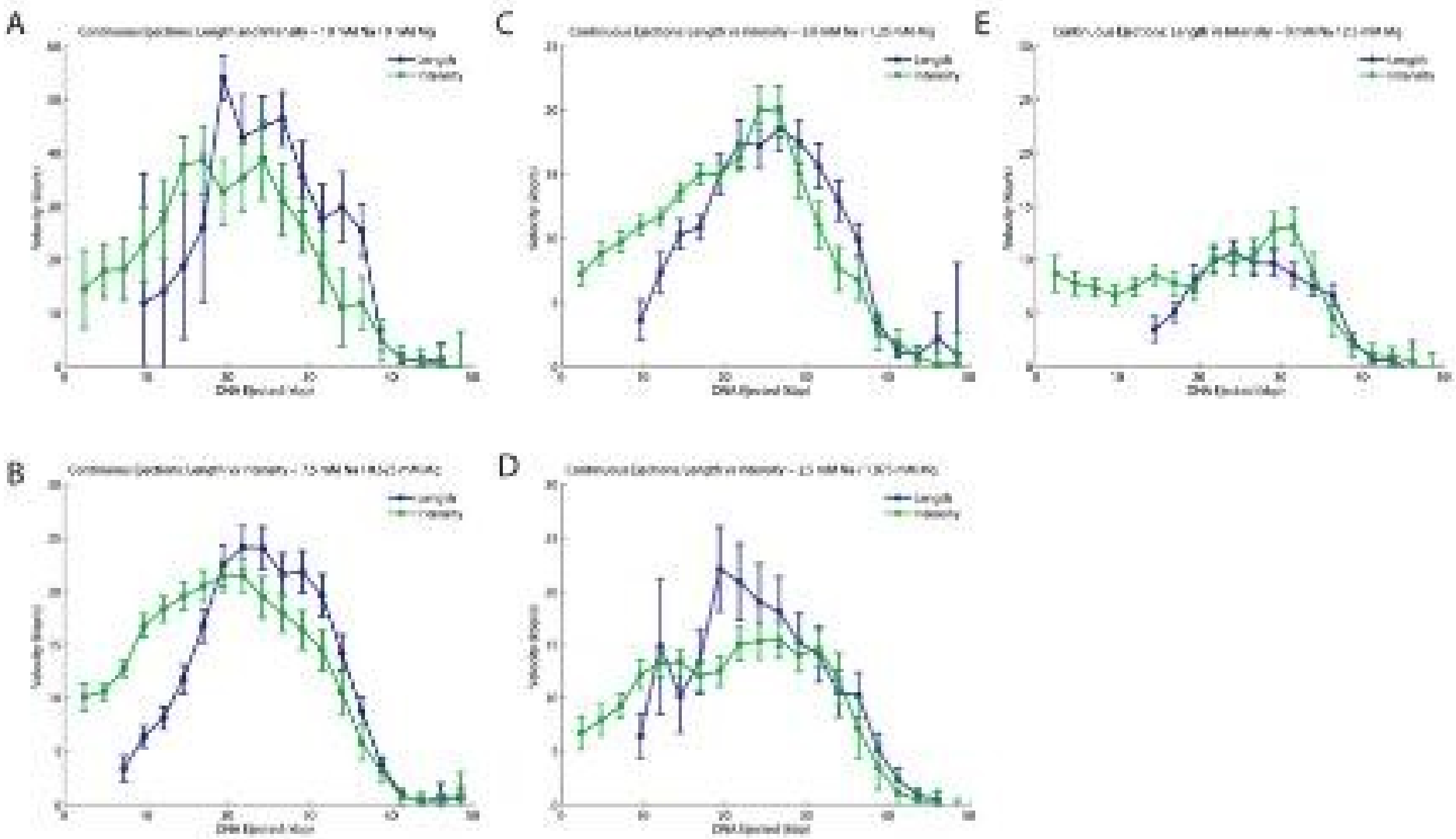

Figure S3. Velocities from the single-phage ejection experiment. Each panel has two classes – green lines denote velocities measured by intensity; blue lines denote velocities measured by length. Intensity measures give a larger estimate for the velocity of DNA ejected at early times because our segmentation algorithm misses the first bits of DNA ejection, and so changes in DNA length appear relatively large at small lengths. Since DNA lengths are hard to directly measure via restriction digest calibration at early times in ejections, we don't measure lengths less than around 5 kbp. The number of trajectories used in each velocity plot is listed in Fig. S1 caption. The errors were determined by standard error. (A) Salt condition: 10 mM Tris, pH 7.4, 10 mM NaCl, 0 mM $MgSO_4$. (B) Salt condition: 10 mM Tris, pH 7.4, 7.5 mM NaCl, 0.625 mM $MgSO_4$. (C) Salt condition: 10 mM Tris, pH 7.4, 5.0 mM NaCl, 1.25 mM $MgSO_4$. (D) Salt condition: 10 mM Tris, pH 7.4, 2.5 mM NaCl, 1.875 mM $MgSO_4$. (E) Salt condition: 10 mM Tris, pH 7.4, 0 mM NaCl, 2.5 mM $MgSO_4$.

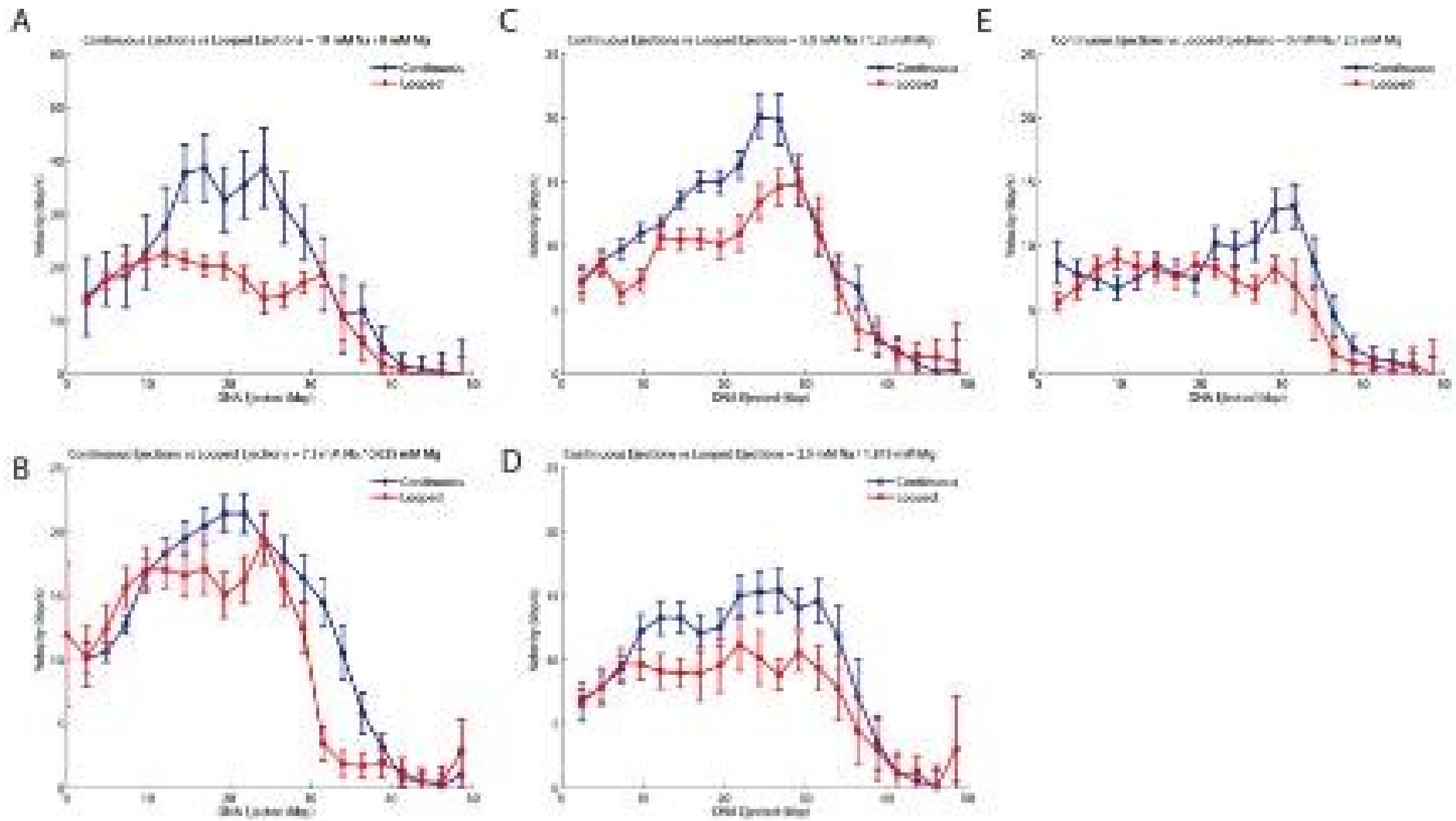

Figure S4. Velocity of DNA at different salt conditions. Red: velocity based on looped trajectories. Blue: velocities based on continuous trajectories. All velocities were determined using intensities; both classes of ejections show similar dynamics. The number of trajectories used in each velocity plot is listed in Fig. S1 caption. The errors were determined by standard error. (A) Salt condition: 10 mM Tris, pH 7.4, 10 mM NaCl, 0 mM MgSO4. (B) Salt condition: 10 mM Tris, pH 7.4, 7.5 mM NaCl, 0.625 mM MgSO4. (C) Salt condition: 10 mM Tris, pH 7.4, 5.0 mM NaCl, 1.25 mM MgSO4. (D) Salt condition: 10 mM Tris, pH 7.4, 2.5 mM NaCl, 1.875 mM MgSO4. (E) Salt condition: 10 mM Tris, pH 7.4, 0 mM NaCl, 2.5 mM MgSO4.

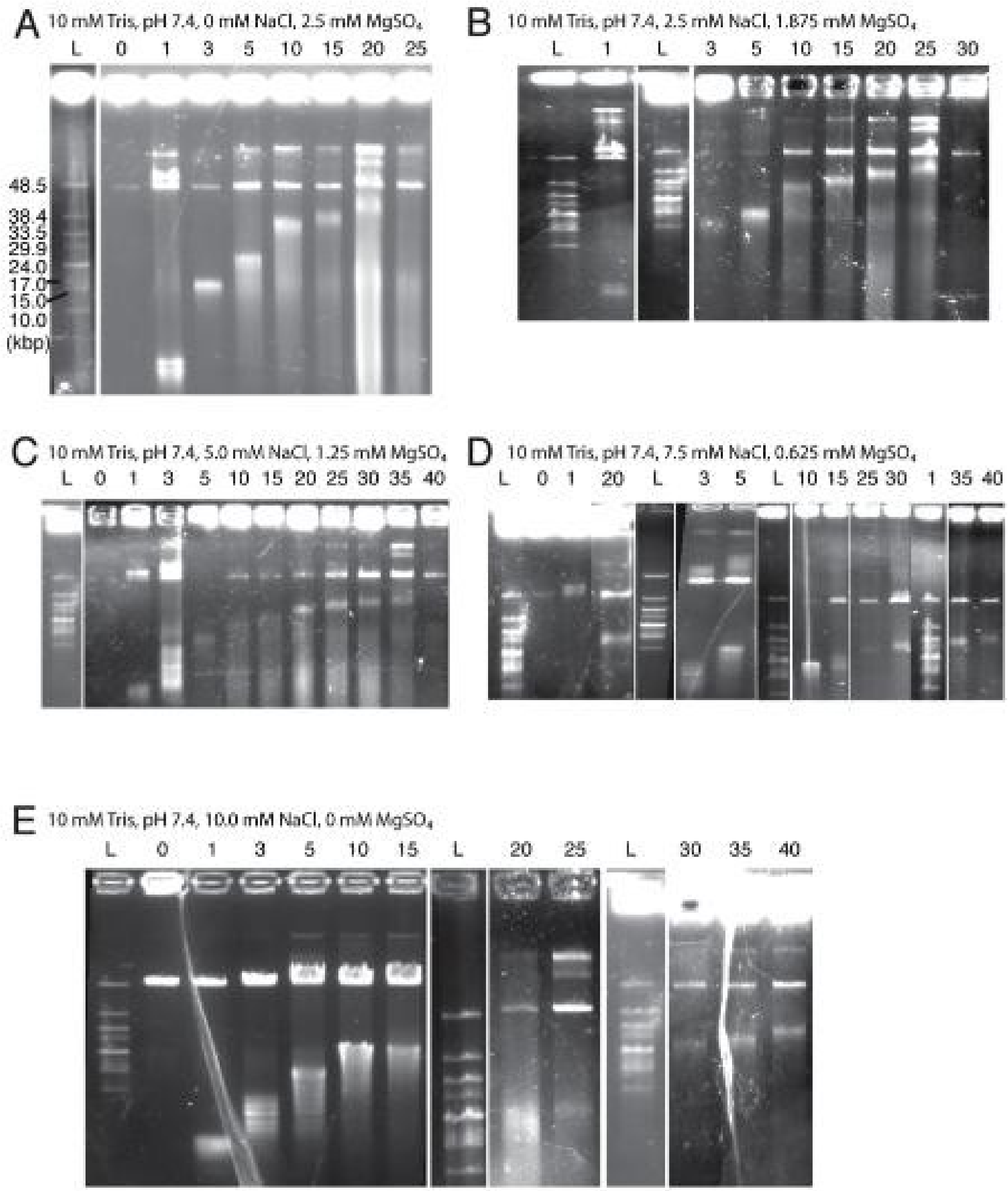

Figure S5. Field-inversion gel electrophoresis of force-dependence assay. A trend of DNA retention in the capsid with increasing external osmotic pressure is clearly visible. (A) Salt condition: 10 mM Tris, pH 7.4, 0 mM NaCl, 2.5 mM MgSO4. (B) Salt condition: 10 mM Tris, pH 7.4, 2.5mM NaCl, 1.875 mM

MgSO4. (C) Salt condition: 10 mM Tris, pH 7.4, 5.0 mM NaCl, 1.25 mM MgSO4. (D) Salt condition: 10 mM Tris, pH 7.4, 7.5 mM NaCl, 0.625 mM MgSO4. (E) Salt condition: 10 mM Tris, pH 7.4, 10 mM NaCl, 0 mM MgSO4.

Additional references: